\documentclass[12pt]{article}
\pdfoutput=1

\usepackage{cite}

\usepackage[nointlimits,reqno]{amsmath}
\usepackage{amssymb}
\usepackage{amsmath}
\usepackage{graphicx}
\usepackage{tabularx}
\usepackage{multirow}
\usepackage{epstopdf}
\usepackage{slashed}
\usepackage{verbatim}
\usepackage[section]{placeins}
\usepackage[space]{grffile}
\numberwithin{equation}{section} 

\usepackage[pdftex,
 pdfproducer=TeXShop,
 pdfcreator=pdflatex]{hyperref}

\usepackage{xspace}
\usepackage{luty}

\usepackage{color}

\newcommand{\bea}{\begin{eqnarray}}
\newcommand{\eea}{\end{eqnarray}}

\begin{document}

\begin{titlepage}

\title{Leptophilic Effective WIMPs}

\author{Spencer Chang$^*$, Ralph Edezhath$^\dagger$, 
Jeffrey Hutchinson$^\dagger$, and Markus Luty$^\dagger$}
   
\address{$^*$Institute of Theoretical Science, University of Oregon\\ 
Eugene, Oregon  97403}

\address{$^\dagger$Physics Department, University of California, Davis\\
Davis, California 95616}

\begin{abstract}
Effective WIMP models are minimal extensions of the standard model 
that explain the relic density of dark matter by the ``WIMP miracle.''
In this paper we consider the phenomenology of effective WIMPs with trilinear
couplings to leptons and a new ``lepton partner'' particle.
The observed relic abundance fixes the strength of the cubic coupling,
so the parameters of the models are defined by the masses of the WIMP and lepton 
partner particles.
This gives a simple parameter space where collider and direct detection
experiments can be compared under well-defined physical minimality assumptions.
The most sensitive collider probe is the search for leptons + MET,
while the most sensitive direct detection channel is scattering 
from nuclei arising from loop diagrams.
Collider and direct detection searches are highly complementary:
colliders give the only meaningful constraint when dark matter is
its own antiparticle, while direct detection is generally more sensitive 
if the dark matter is not its own antiparticle.
\end{abstract}
\end{titlepage}

\section{Introduction}
\label{sec:intro}
The observation of dark matter is an unambiguous discovery of physics beyond
the standard model.
The simplest and most compelling explanation for dark matter is a stable
thermal relic, a weakly interacting massive particle (WIMP).
Assuming that it freezes out by annihilation to standard model particles
via dimensionless order-1 couplings, the correct relic 
density is obtained for dark matter masses of order 100~GeV to 1~TeV.
This ``WIMP miracle'' means that dark matter can potentially be directly
produced and studied at high-energy collider experiments such as the LHC.
At the same time, this coupling allows for direct detection of astrophysical
dark matter through its interactions with ordinary matter. 
It is one of the most 
ambitious dreams
of particle physics to be able to
discover dark matter in both types of experiment, and make a direct comparison
to properties of dark matter inferred from cosmology and astrophysics.

There are many well-motivated models of physics beyond the standard model
that have a WIMP candidate (for example, the minimal supersymmetric standard
model), but comparing collider and direct detection sensitivity is generally 
model-dependent due to the large number of 
independent parameters
in these theories.
In a previous paper \cite{Chang:2013oia}, we proposed a class of minimal models
where it is assumed that the only particles beyond the standard model that
are important for dark matter phenomenology consist of the WIMP and a
standard model ``partner'' particle with the same gauge quantum numbers as
one of the standard model particles.
This allows a cubic coupling of the form
\beq\eql{cubic}
\De \scr{L} \sim \la (\text{SM})  ( \widetilde{\text{SM}}) (\text{DM}),
\eeq
where DM denotes the dark matter particle and 
SM ($\widetilde{\text{SM}}$) denote the standard model field and its
partner, respectively.
The coupling $\la$ is fixed by requiring the correct relic abundance, so 
the only parameters in this model are the masses of the WIMP and the
standard model partner particle, once the spin and $CP$ properties
of the dark matter is fixed.
This gives a well-defined 
and complete
effective theory for dark matter phenomenology,
motivating the name ``effective WIMP'' for this class of models.
In particular, it allows direct comparison of the collider and direct detection searches
for dark matter under well-defined physical minimality assumptions.

\Ref{Chang:2013oia} analyzed the phenomenology in the case where the standard
model particle and its partner are colored.
In this case collider and direct detection experiments were found
to be highly complementary.
Similar models were also studied 
in \Refs{An:2013xka, Bai:2013iqa,DiFranzo:2013vra,Papucci:2014iwa}.
The major difference in our work is that we focus on the parameter
space where the WIMP has the correct thermal relic abundance.  
This has important implications for the global picture.
For example, \Ref{Chang:2013oia} found that the coupling $\la$ is enhanced
in some regions of parameter space, leading to increased sensitivity for
collider and direct detection searches.
On the other hand, the relic abundance constraint eliminates regions
of parameter space where indirect detection limits are important.
There has also been other approaches to simplified dark matter such as $s$-channel 
mediators \cite{Buchmueller:2013dya,An:2012ue,Shoemaker:2011vi}
and dark matter with electroweak interactions \cite{Cirelli:2005uq, Cheung:2013dua}.

In this paper, we consider effective WIMP models where the standard model
particle is a lepton ($e, \mu, \tau$ or their corresponding neutrinos).
In this case, the collider constraints come from the pair production of
lepton partners followed by their decay into a WIMP plus a lepton or
neutrino, and so the collider constraints are independent of the value
of $\la$.
Once again, we find that collider and direct detection experiments are
highly complementary.
The direct detection constraints depend on $\la$, with the dominant interactions coming
from loop diagrams to a photon.  
The type of interactions allowed depend on whether the dark matter is its own 
antiparticle, which strongly affects the scattering rate.  
If the dark matter is its own antiparticle, the only meaningful constraints
come from collider experiments, while in the opposite case direct
detection experiments are generally more sensitive.
There is a significant region of parameter space where both future collider
and direct detection experiments can see a signal.

Imposing the relic abundance constraints has an important effect on the
direct detection phenomenology.
In particular, these constraints extend into the regime of large
lepton partner masses because the coupling $\la$ becomes large in this
region to get the correct relic abundance.
Imposing the relic abundance constraint also implies that constraints
from indirect detection are not important.

The models that we consider are listed in Table~1.
We consider couplings to left-handed lepton doublets only,
since the phenomenology of right-handed leptons is essentially
a subset of this case.  Note that this fixes the production rate at colliders which leads to slightly stronger constraints than the right-handed lepton partner case.
We consider both the cases where the dark matter couples to only the first two
generations of leptons, as well as the case where it couples to all generations.
We find that the differences between them are small, as one might expect.

\begin{table}
\begin{center}
\begin{tabularx}{.85\textwidth}{|p{4cm}|p{4cm}|X|}
\hline
\multicolumn{2}{|c|}{Particles} & \multirow{2}{.2\textwidth}{\parbox{4cm}{\hspace{.5in} $\mathcal{L}_{\text{int}}$}}\\
\cline{1-2}
  Dark matter $\chi$ & Lepton partner $L$ &     \\ 
\hline
\hline
 Majorana fermion& Complex scalar&  $\la(\chi \ell)L^*+\text{h.c.} $\\ 
\hline
Dirac fermion & Complex scalar & $\la(\chi \ell)L^*+\text{h.c.}$ \\
\hline
Real scalar & Dirac fermion&  $\la (L^c \ell)\chi+\text{h.c.}$  \\
\hline
Complex scalar & Dirac fermion & $\la (L^c \ell)\chi+\text{h.c.}$   \\
\hline

\end{tabularx}
\begin{minipage}{5.5in}
\caption{\small
Summary of the models considered in this paper.
Spinors are written in 2-component notation.
Here $\ell$ is the left-handed lepton doublet of the standard model,
$L$ is the lepton partner field, and $\chi$ is the dark matter field.
\label{table:models}}
\end{minipage}
\end{center}
\end{table}

\section{General Features\label{sec:general}}

\subsection{Relic Abundance}
\begin{figure} [h]
\centering  \includegraphics[scale=.7]{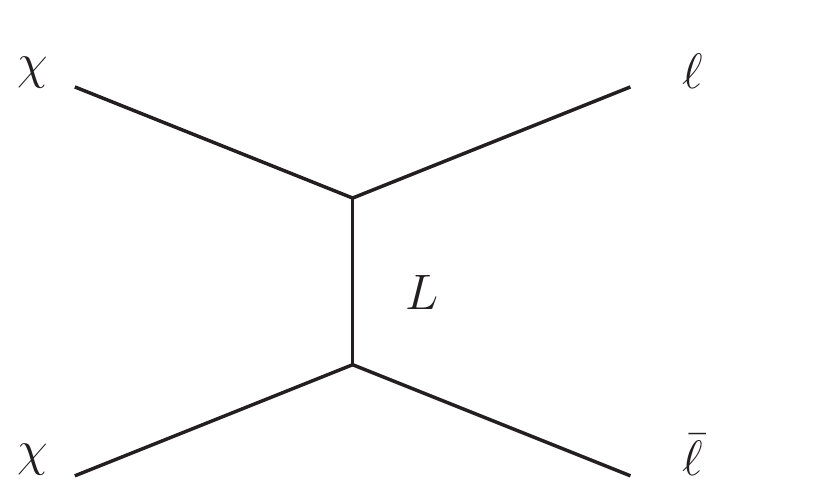}
\begin{minipage}{5.5in}
\caption{\small 
Feynman diagram contributing to dark matter
freeze-out
\label{fig:annihilation}}
\end{minipage}
\end{figure}

The relic abundance of non-baryonic matter is  accurately determined
by cosmological constraints
to be $\Omega_{\chi} h^2 = 0.1199 \pm 0.0027$ \cite{Ade:2013lta}.  
We assume that the dark matter is entirely composed of the WIMP in our model,
denoted henceforth by $\chi$.
Under the assumption that $\chi$ particles were in thermal equilibrium in
the early universe, its present relic density is determined by freeze-out
from the annihilation process $\chi\bar\chi \to \ell \bar{\ell}$ shown in 
Fig.~\ref{fig:annihilation}.
The relic abundance is determined by the thermally averaged annihilation
cross section $\avg{\si(\chi\bar\chi \to \bar{\ell}\ell) v}$
at temperatures $T_f \sim m_\chi / 25$.
The dark matter velocity during annihilation is $v^2 \sim 0.1$, so we can expand
\beq
\si(\chi\bar\chi \to \bar{\ell}\ell) v = a + b v^2 + O(v^4).
\eeq
Approximate formulas for the relic density in terms of these
parameters are given in Appendix A.
The coefficients $a$ and $b$ represent $s$-wave
and $p$-wave contributions,
and are computed in each model below.

\subsection{Collider Limits\label{sec:colliderlimits}}
The dark matter can be produced at colliders by Drell-Yan production
via an intermediate $W$, $Z$, or $\ga$.
The relic abundance constraint means that the coupling $\la$ is always
sufficiently large that the lepton partner decays promptly.
The collider phenomenology is therefore independent of $\la$, and depends
only on the masses and the spin of the new particles.
The charged lepton partners decay to $\ell\, \chi$, while the neutral lepton
partners decay invisibly to $\nu\, \chi$,
so there are signal events with 0, 1, or 2 charged leptons.
This paper focuses on the dilepton channel, but we make some brief remarks
about the other channels below.

The totally invisible channel can be probed by monojet searches, but requiring a jet from initial state radiation leads to a small rate with weak constraints.
The 1 lepton $+$ MET channel has been searched for in 
the context of $W'$ models \cite{Chatrchyan:2013lga}.
We have checked that the event rates for lepton partner production are too
small to provide the most important collider constraints in this channel, even in the case
of fermionic lepton partners which have the largest rate.

We therefore turn to the dilepton plus MET channel.
This has been searched for by
both ATLAS \cite{ATLAS-CONF-2013-049} and CMS \cite{CMS-PAS-SUS-13-006}, which very helpfully have cross section limits as a function of the particle masses.
We use the most recent ATLAS results for this paper.
The event rates were calculated at parton level using 
{\tt MadGraph5} v1.5.11 \cite{Alwall:2011uj}.
The {\tt MadGraph} model files were generated using {\tt FeynRules} v1.6.0 \cite{Christensen:2008py}.

\begin{figure}[h]
\centering \includegraphics[scale=0.6]{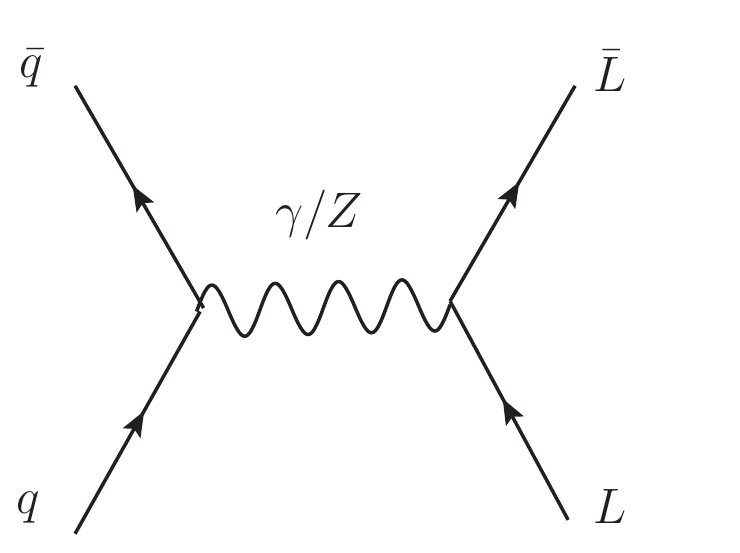}
\begin{minipage}{5.5in}
\caption{\small 
Feynman diagram contributing to leptons + MET signal at colliders.
\label{fig:collider}}
\end{minipage}
\end{figure}

\begin{figure}[t]
\centering 
\begin{minipage}{0.45\textwidth}
\centering \includegraphics[scale=0.6]{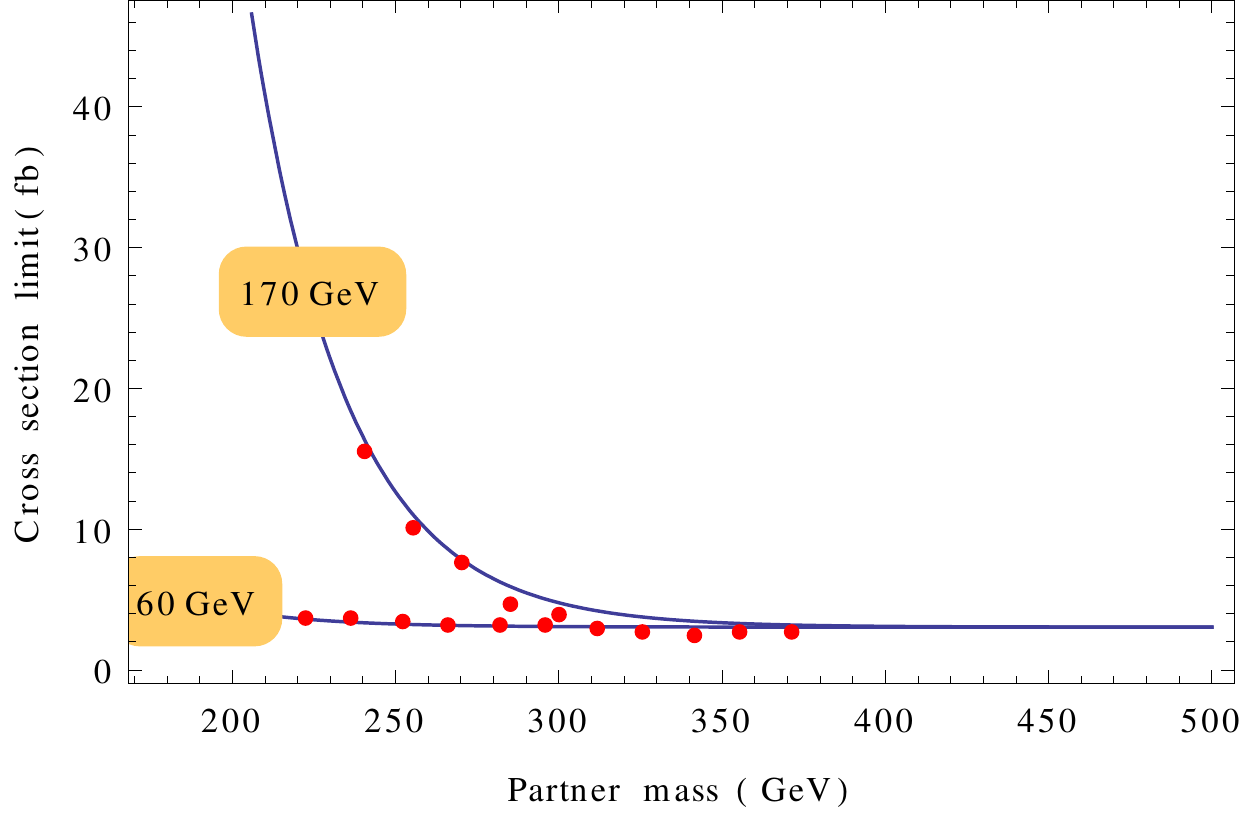}
\caption{\footnotesize{Extrapolation of limits for 2 lepton $+$ MET search
as a function of $m_L$ for fixed $m_\chi$,
compared with ATLAS reported limits.
\label{fig:fixedmx}}
}
\end{minipage} \hspace{12mm}
\begin{minipage}{0.45\textwidth}
\centering \includegraphics[scale=0.6]{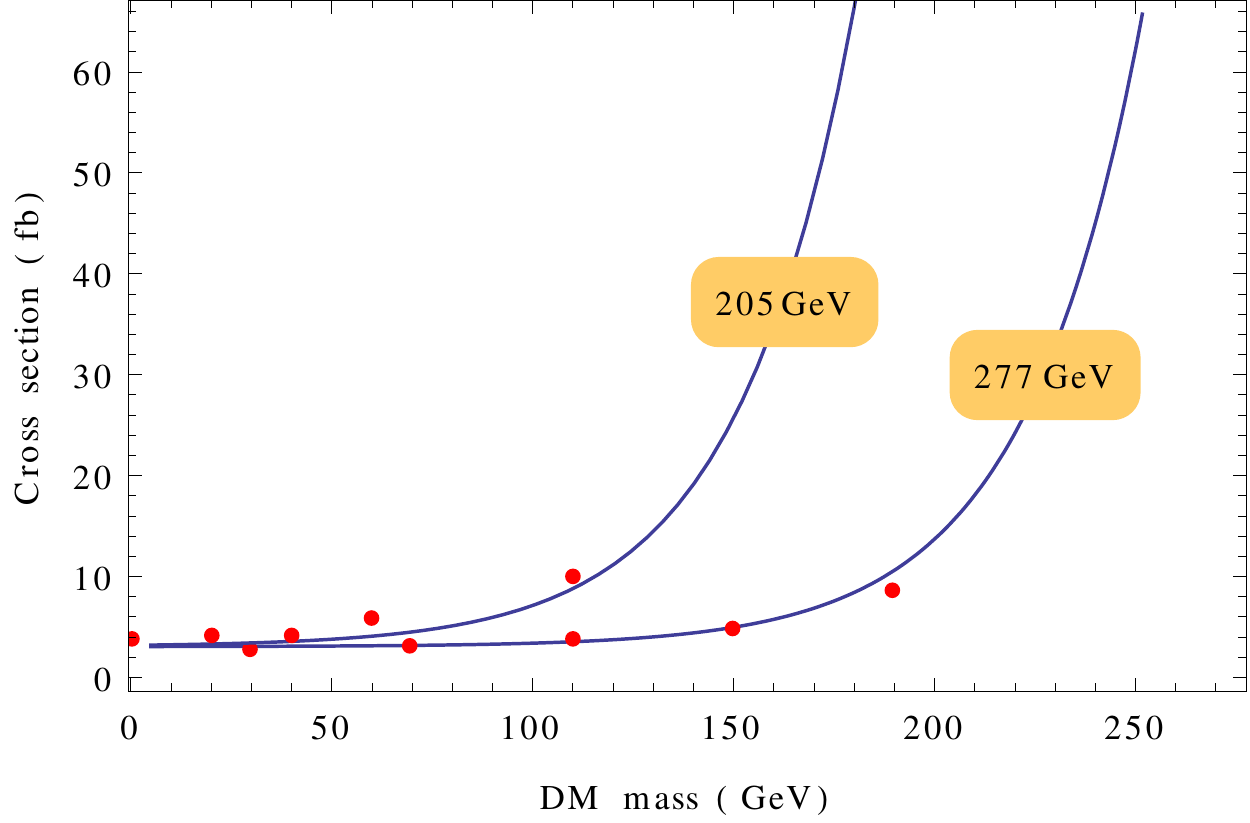}
\caption{\footnotesize 
Extrapolation of limits for 2 lepton $+$ MET search
as a function of $m_\chi$ for fixed $m_L$,
compared with ATLAS reported limits.
\label{fig:fixedml}}
\end{minipage}
\end{figure}

The ATLAS search was for sleptons, which have a lower cross
section than fermionic lepton partners with the same mass.
They report limits on $\si \times \text{BR}$ only for slepton
masses up to $\sim 350\GeV$ and dark matter masses up to $\sim 200\GeV$,
but their search should be sensitive to fermionic lepton partners beyond
this range.  
We extrapolate the ATLAS search in the following way.
The maximum allowed $\si \times \text{BR}$ is expected to depend
mainly on $m_L - m_\chi$, which controls the $p_T$ of the lepton
and the MET.
The bound gets weaker for small values of $m_L - m_\chi$,
so we fitted the reported ATLAS results to the function
\beq
(\si \times \text{BR})_\text{limit} = a + b \gap e^{-c (m_L-m_\chi)}.
\eeq  
This works well over the range reported by ATLAS, as can be seen in
Figs.~\ref{fig:fixedmx} and \ref{fig:fixedml}.  
The CMS search has limits for a larger mass range, 
but we find it has weaker limits than the extrapolated ATLAS limits,
so we use the latter in our plots.

\FloatBarrier

\subsection{Direct Detection\label{sec:directdetection}}

For dark matter interacting with leptons, there are important modifications to
the sensitivity of direct detection experiments.
To couple to the nucleus, interactions occur at loop level through induced electromagnetic 
form factors of the dark matter \cite{Pospelov:2000bq}, 
such as a charge radius or magnetic moment.  
There have also been analyses of the possibility of detecting dark matter scattering 
from electrons in the atoms \cite{Bernabei:2007gr, Kopp:2009et, Essig:2011nj} with a 
published limit using XENON10 data \cite{Essig:2012yx}.  
However, these processes are more important for dark matter masses below a GeV, where 
the dark matter has insufficient kinetic energy to give detectable ($\sim\text{keV}$)
nuclear recoil energies.

\begin{figure}[h!]
\centering 
\centering \includegraphics[scale=0.6]{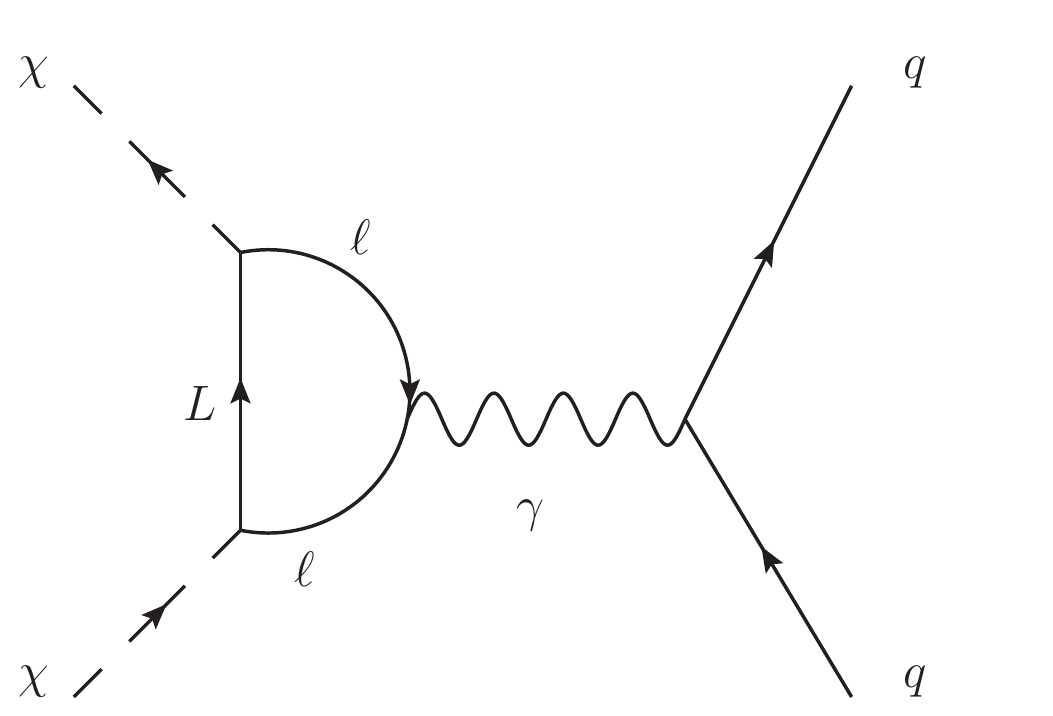}
\caption{\footnotesize{Schematic 1-loop contributions for direct detection.  There are also diagrams where the lepton-partner (L) and lepton ($\ell$) are interchanged.\vspace{5mm}}
\label{fig:OneLoopDD}}
\end{figure}

In this paper, we focus on dark matter heavier than a GeV, so from now on we will consider 
only scattering with the nucleus.   
The leading WIMP-nucleus interactions arise
at 1-loop level through diagrams of the form Fig.~\ref{fig:OneLoopDD}. 
This gives rise to interactions to the photon through a charge radius operator
\beq
\scr{L}_\text{charge radius} =  \Bigg\{ \begin{matrix} 2 b_\chi\, \partial_\mu \chi^* \partial_\nu \chi\, F^{\mu \nu} & \text{ (complex scalar DM)} \\ b_\chi\, \bar\chi \gamma_\nu \chi\, \partial_{\mu} F^{\mu \nu}& \text{ (Dirac fermion DM)} \end{matrix}
\eeq
and a magnetic moment operator for the Dirac fermion
\bea
\scr{L}_\text{magnetic moment} = 
&& \frac{\mu_\chi}{2} \bar\chi^* \sigma_{\mu\nu} \chi  F^{\mu \nu} \text{ (Dirac fermion DM)}.
\eea
See sections \ref{diracdd} and \ref{complexdd} in the appendix
for detailed results of our calculation of these coefficients.  
As a check, we confirmed that our formulas agree for the Dirac fermion case with 
\cite{Agrawal:2011ze, Batell:2013zwa} in the relevant limits.  
Of the two contributions to the scattering of a Dirac fermion, the charge radius 
is typically larger because the charged radius strength has a logarithmic enhancing factor of
$\ln(m_L^2/m_\ell^2) \sim 10$.
The scattering cross section from this operator is given by 
\beq\eql{chargeradius}
\frac{d\sigma_{b_\chi}}{dE_R} = \frac{m_N}{2\pi v^2}Z^2 e^2 b_\chi^2 F^2[E_R]
\eeq
where $F[E_R]$ is the nucleus charge form factor.  
This is a spin-independent interaction, so we use the current best limits from LUX's 
$85.3$ day run \cite{Akerib:2013tjd}. 
We also use projected sensitivities for XENON1T taken from {\tt DMtools} 
\cite{DMtoolsProjections}.  
 
\begin{figure}[h!]
\centering \includegraphics[scale=0.6]{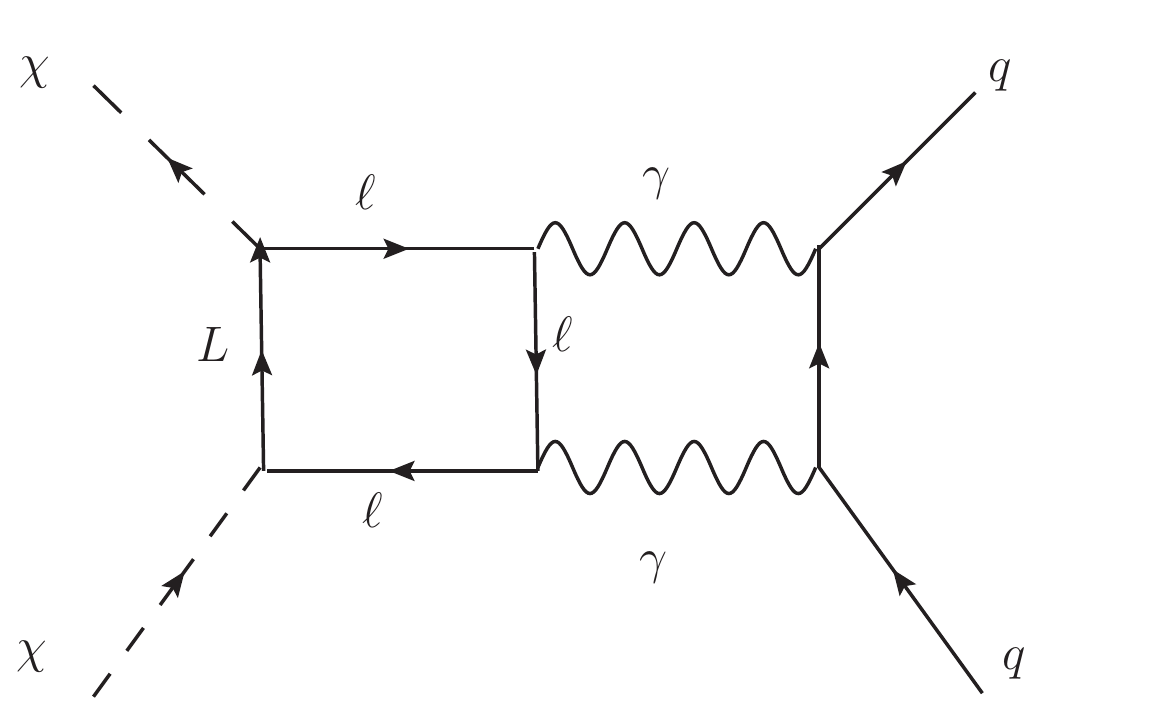}
\caption{\footnotesize 
Schematic 2-loop contributions to direct detection scattering. 
 There are also diagrams where the lepton-partner (L) and lepton ($\ell$) are interchanged and where the photons are crossed.  \label{fig:TwoLoopDD}}

\end{figure}

When the dark matter is its own antiparticle, symmetry requires the charge radius and magnetic moment to vanish.  
For real scalar dark matter, there is no magnetic moment because it is
spin-0, and the charge radius operator vanishes due to antisymmetry of the 
field strength tensor.
Therefore there is no 1-loop contribution in this case.
The leading interaction with nuclei arises from 2-loop interactions such as the 
one shown in Fig.~\ref{fig:TwoLoopDD}.
The additional suppression reduces the WIMP-nucleus cross section to well below 
current limits. 

For Majorana dark matter, the vector and tensor operator are zero, so the dipole moment and the charge radius
vanish.
The leading interaction comes from an anapole moment operator of the form
\beq
\bar{\chi}\gamma_5 \gamma_\mu \chi\; \partial_\nu F^{\mu \nu},
\eeq
which also leads to a cross section below direct detection limits, except for the case where the dark matter and lepton partner are extremely degenerate, as discussed recently in \cite{Kopp:2014tsa}.   
See  \S\ref{realscalardd} and \ref{Majoranadd} in the appendix for more details on the direct detection for Majorana and real scalar dark matter.


\FloatBarrier

As will be shown later, direct detection experiments provide the most 
stringent limits for Dirac fermion or complex scalar dark matter. 
These limits extend to extremely high masses in two regions of parameter space: 
near $m_L \sim m_\chi$  and $m_L \gg m_\chi$.  
In the first region the particles are nearly degenerate, and 
there is an enhancement due to the possibility of the virtual partner particle 
going (nearly) on-shell in Fig.~\ref{fig:OneLoopDD}.  
This can be seen in the charge radius \Eq{diracbx} which diverges in the limit 
$m_\chi \rightarrow m_L$.  
On the other hand, for the $m_L \gg m_\chi$ region, the direct detection 
constraint remains strong because the coupling $\la$ is required to become
larger to obtain the observed relic abundance.  
Fixing the dark matter mass, the annihilation cross section requires the interaction strength to scale as 
\beq
\lambda^4 \sim 4\pi \sigma_\text{ann} \frac{m_L^4}{m_\chi^2}.
\eeq
At large $m_L$, the charge radius is given by 
\bea
b_\chi \sim \frac{\lambda^2 e}{16\pi^2 m_L^2}\left[1+\frac{2}{3}\ln{\frac{m_\ell^2}{m_L^2}}\right] \sim  \frac{e \sqrt{4\pi \sigma_\text{ann}}}{16\pi^2 m_\chi}\left[1+\frac{2}{3}\ln{\frac{m_\ell^2}{m_L^2}}\right] 
\eea
Thus, aside from the logarithm, the increase in $m_L$ does not change the charge radius or the scattering cross section, leading to an asymptotic limit on $m_\chi$. 

\begin{table}
\begin{center}
\begin{tabular}{|c|c|c|c|}
\hline
  \multicolumn{2}{|c|}{Model} & \multirow{2}{*}{{Relic Abundance}} &  \multirow{2}{*}{\parbox{3cm}{Direct Detection}} \\
\cline{1-2}
  $\chi$ & L & &     \\ 
\hline
 Majorana fermion& Complex scalar& \begin{tabular}[x]{@{}c@{}} $a \sim m_\ell^2$\\$ \la \sim 0.5-3$\end{tabular}&  \begin{tabular}[x]{@{}c@{}} Anapole Scattering \\ \end{tabular}    \\
\hline
Dirac fermion & Complex scalar &  $\la \sim 0.2-1$ &  \begin{tabular}[x]{@{}c@{}} One loop charge radius\\ $\si_\text{SI} \overset{m_L\gg m_\chi}{\sim}  \frac{m_p^2}{m_\chi^2}\si_\text{ann} $\end{tabular}\\
\hline
Real scalar & Dirac fermion& \begin{tabular}[x]{@{}c@{}} $a,b \sim  m_\ell^2 $\\$ \la \sim 1-7$\end{tabular} &  \begin{tabular}[x]{@{}c@{}} Two loop scattering\\ through two photons \\ \end{tabular}\\
\hline
Complex scalar & Dirac fermion &  \begin{tabular}[x]{@{}c@{}} $a \sim m_\ell^2 $\\$ \la \sim 0.5-3$\end{tabular} &   \begin{tabular}[x]{@{}c@{}}One loop charge radius\\  $\si_\text{SI} \overset{m_L\gg m_\chi}{\sim}  \frac{m_p^2}{m_\chi^2}\si_\text{ann} $\end{tabular}\\
\hline

\end{tabular}
\begin{minipage}{5.5in}
\caption{Overview of results for relic abundance and direct
detection for the various models. \label{table:results}}
\end{minipage}
\end{center}
\end{table}

\subsection{Indirect Detection \label{sec:indirectdetection}}
In these models, the dark matter in our galaxy will annihilate into charged leptons and neutrinos with a cross section bounded by that required by the relic abundance.  So all of these annihilation cross sections are bounded $\langle \si(\chi\bar\chi \to \bar{\ell}\ell) v\rangle \le 3\times10^{-26} \text{ cm}^3/\text{s}$.  The only indirect detection constraints that are sensitive to such a cross section is annihilation into tau leptons.  These are constrained by the resultant gamma ray production from the tau decays that could have been seen by Fermi observations of Milky Way satellites \cite{GeringerSameth:2011iw, Ackermann:2011w a}.  These set an upper limit on the dark matter mass, which is sensitive to systematic uncertainties of the dark matter distribution of these objects.  The analysis \cite{GeringerSameth:2011iw} shows that this uncertainty allows the limit to range from $m_\chi < 13-80$ GeV.  So there is a limit for models coupling to the third generation leptons but there are significant uncertainties, so we choose to omit it from our later plots.

\subsection{Muon Anomalous Magnetic Moment\label{sec:muong2}}
The anomalous magnetic moment of the muon $a_\mu$ receives corrections from loops
of $\chi$ and $L$ particles in our models.
There is nominally a $\sim 3\si$ discrepancy between the measured value of $a_\mu$
and the value predicted in the standard model (for a review, see \cite{Blum:2013xva}).
However, there are large theoretical uncertainties, particular in the
light-by-light scattering which can only be estimated from strong interaction models \cite{Blum:2013xva}.
Our point of view is that corrections to $a_\mu$ that make the discrepancy
significantly worse can be used to constrain new physics, but we do not 
believe that regions that reduce the discrepancy are favored.

The contributions to the photon-muon-muon amplitude are parameterized as
\beq
\mathcal{M} = ie\bar{u}\left ( \gamma^\lambda + (a_\mu^{SM} + \delta a_\mu ) \frac{i \sigma^{\lambda \beta}q_\beta}{2 m_\mu} \right ) u \epsilon _\lambda
\eeq
We can read off the contributions from the calculations in supersymmetric
models, computed in \Ref{Martin:2001st}.  
For fermionic dark matter, $\chi$ is identified as a neutralino, 
and $L$ is identified as a slepton, and we obtain
\beq
\delta a_{\mu, fermion} = - \frac{\lambda^2 m_\mu^2}{192 \pi^2 m_L^2} \frac{2}{(1-r_L)^4} \left ( 1-6r_L+3r_L^2+2r_L^3 -6r_L \ln(r_L) \right )
\eeq
where $r_L = m_\chi^2 / m_L^2$.  
Note that this contribution is in the opposite direction of the observed discrepancy.  
For scalar dark matter, $\chi$ is identified as a sneutrino, and $L$ is identified 
as a chargino, and we obtain
\beq
\delta a_{\mu, scalar} =  \frac{\lambda^2 m_\mu^2}{192 \pi^2 m_\chi^2} \frac{2}{(1-r_L^{-1})^4} \left ( 2+3r_L^{-1}-6r_L^{-2}+r_L^{-3} - 6r_L^{-1} \ln(r_L) \right )
\eeq
This contribution is in the same sign as the observed discrepancy.
For both scalar and fermion dark matter, these contributions are comparable to
the observed discrepancy only for very light dark matter, $\lsim 10\GeV$.
We focus on heavier dark matter, and therefore do not consider these
constraints when presenting our results.

\section{Results}

\subsection{Dirac Dark Matter}
\begin{figure}[h!] 
\begin{center}
\includegraphics[scale=.5]{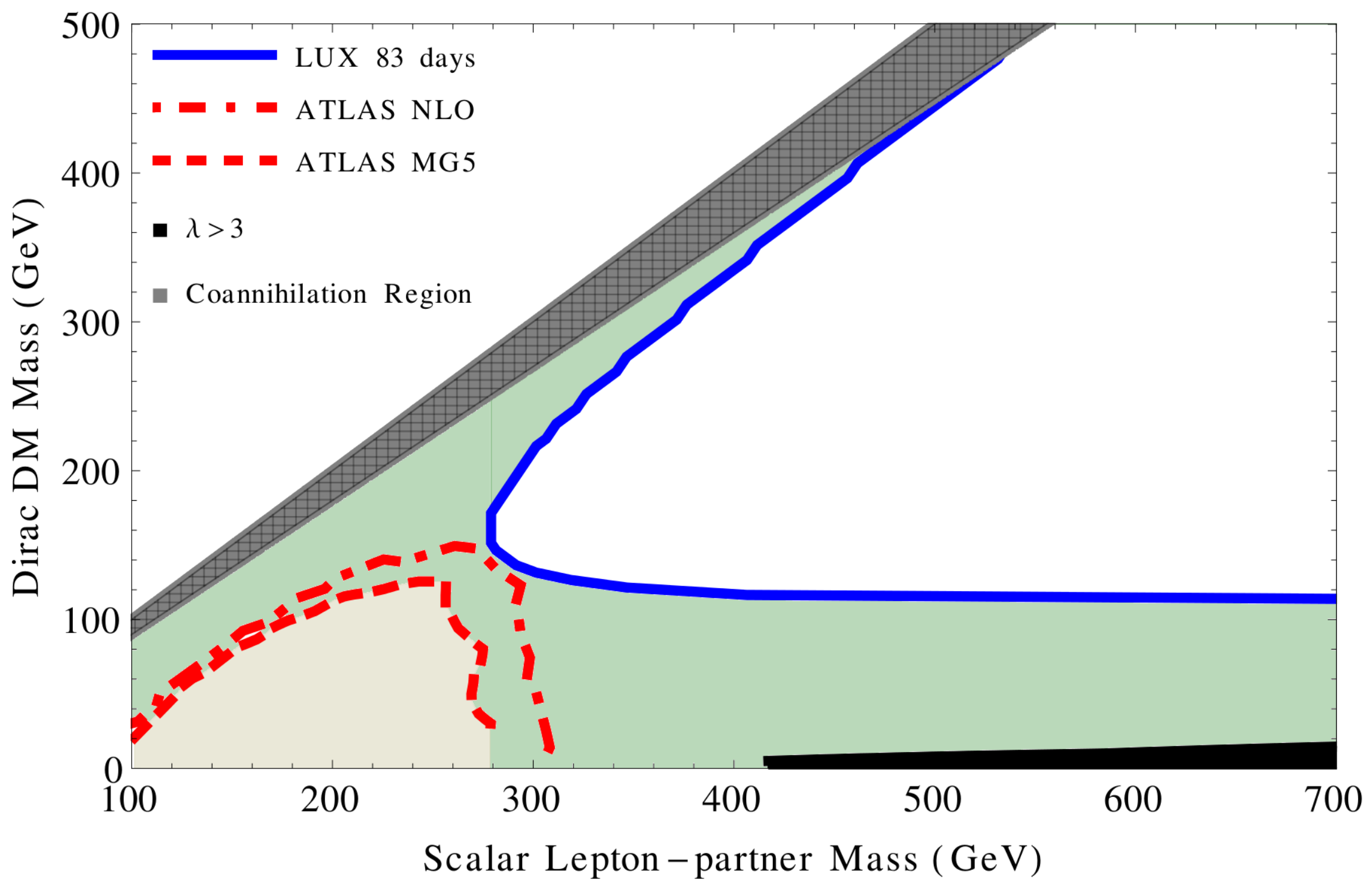}
\caption{\small 
Limits on Dirac dark matter coupling to first 2 generations of leptons only.  }
\label{M1F4}
\end{center}
\end{figure}

\begin{figure}[h!]  
\begin{center}
\includegraphics[scale=.5]{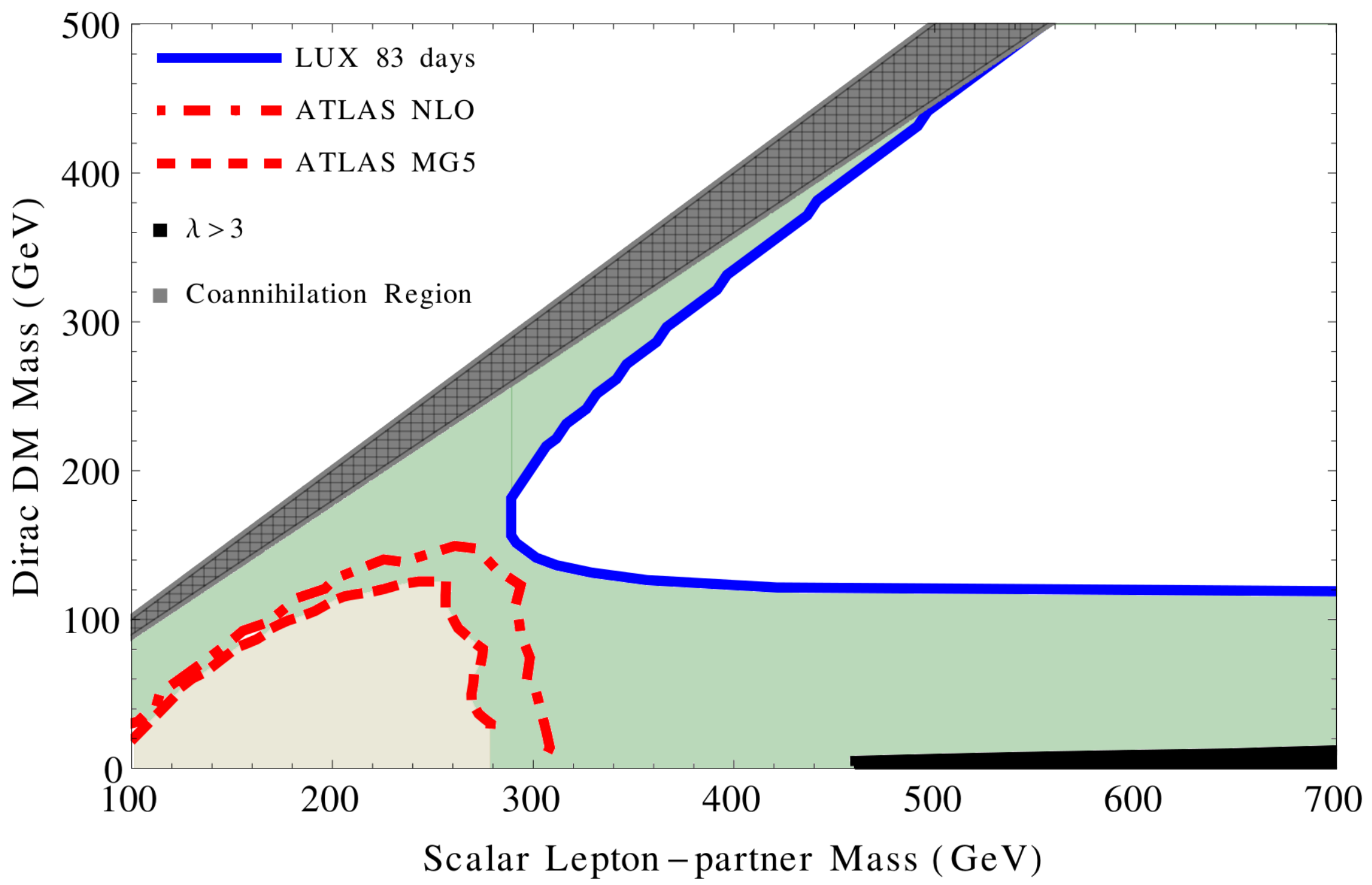}
\caption{\small 
Limits on Dirac dark matter coupling to all 3 generations of leptons. }
\label{M1F1}
\end{center}
\end{figure}

The collider and direct detection constraints for Dirac dark matter are shown
in Fig.~\ref{M1F4} for the coupling to the first two generations of leptons,
and in Fig.~\ref{M1F1} for the coupling to all three generations.
The results of these two cases
are very similar, and probably within the errors of the bounds. 
A collider limit on $\tau$ partners would give an additional probe of
the model that couples to all generations of leptons. 

The dominant direct detection constraint comes from WIMP-nucleus scattering.
The contribution from the charge radius operator \Eq{chargeradius}
dominates over the contribution from the dipole operator \Eq{diracdipole}.
This is a spin-independent interaction, so we use the limits from LUX.
The LUX limits are strong in the degenerate region $m_\chi \simeq m_L$.
But this is also the region where co-annihilation effects are important, 
so our results are not reliable.  As discussed earlier, at large $m_L$ there is a asymptotic limit on $m_\chi$ due to the
increase in the interaction strength as mandated by the relic abundance constraint. 
Although we focus on heavy dark matter, we note that dark matter masses
below approximately 8~GeV are not ruled out by LUX because the smaller
energy deposit results in reduced sensitivity and there should then be an allowed region for LUX at light enough dark matter mass.    

The projected limits for XENON1T extend to dark matter masses up to approximately
650~GeV, covering essentially the entire parameter space shown in 
Figs.~\ref{M1F4} and \ref{M1F1}.  Thus, if XENON1T sees no excess, then we should not expect to see any lepton partner signals in the next run of the LHC.    However, there is still a large viable region around the elbow of the LUX limit where we could have future correlated signals at XENON1T and LHC.

\FloatBarrier

\subsection{Complex Scalar Dark Matter}
\begin{figure}[h!] 
\centering \includegraphics[scale=.5]{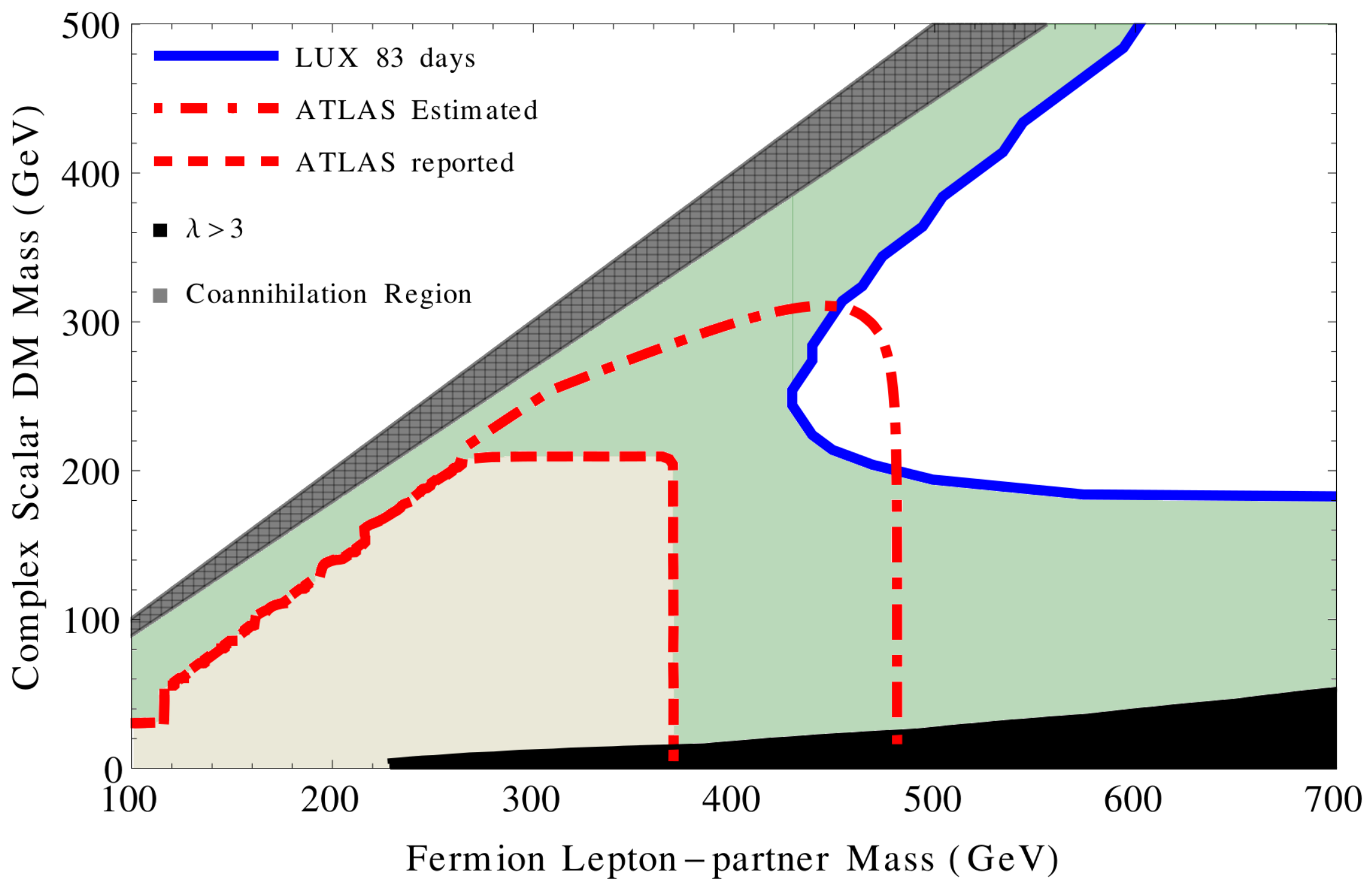}
\begin{minipage}{5.5in}
\caption{\small 
Limits on complex scalar dark matter coupling to first 2 generations of leptons.  }
\label{M2F4}
\end{minipage}
\end{figure}

\begin{figure}[t!] 
\centering \includegraphics[scale=.5]{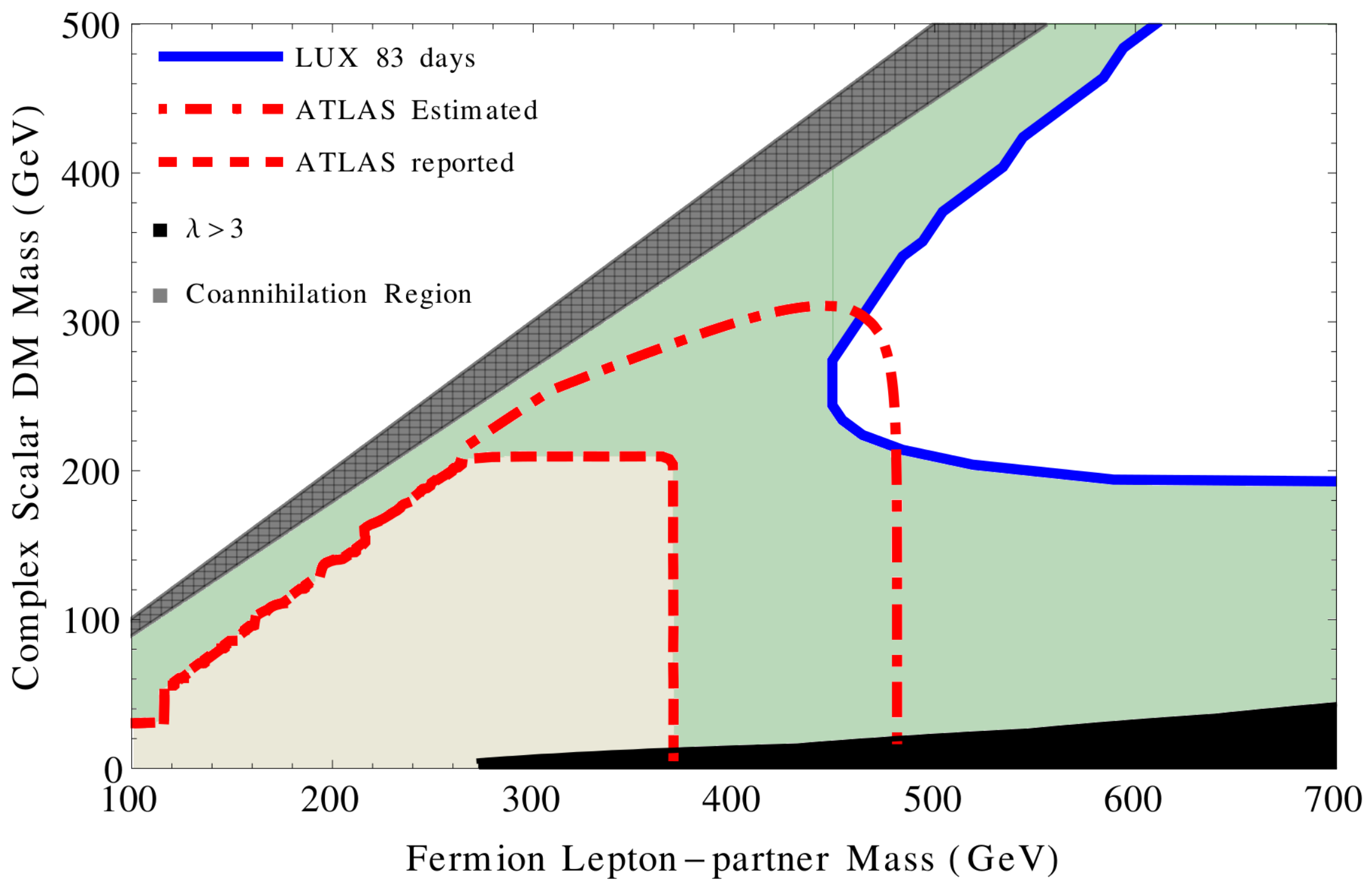}
\begin{minipage}{5.5in}
\caption{\small 
Limits on complex scalar dark matter coupling to all 3 generations of leptons. }
\label{M2F1}
\end{minipage}
\end{figure}

The collider and direct detection constraints for complex scalar
dark matter are shown
in Fig.~\ref{M2F4} for the coupling to the first two generations of leptons,
and in Fig.~\ref{M2F1} for the coupling to all three generations.
The results of these two cases
are very similar, and probably within the errors of the bounds. The collider limits are stronger than for scalar lepton partners
because of the larger production cross section.
As discussed in the main text, the ATLAS dilepton search does not report
bounds for the full range of sensitivity.
We show both the reported and extrapolated bounds and it is clear that there should be sensitivity to higher partner masses.  In fact, our
extrapolation shows that it could be stronger than the LUX direct detection bounds.  

As in the Dirac dark matter case, the LUX bounds are strong both in
the degenerate region and the region of large $m_L$.
Again, dark matter masses below 8~GeV are not constrained.
The projected XENON1T sensitivity covers the entire parameter space
shown in Figs.~\ref{M2F4} and \ref{M2F1}, and extend up to dark matter
masses of approximately 1~TeV.  Once more there is viable parameter space which will be covered by future LHC data and direct detection experiments.

This model can account for the observed discrepancy in the muon anomalous moment
for $2.2\GeV < m_\chi < 7.3\GeV$ for dark matter coupling to the first 2 generations
of leptons, and $1.8 \GeV < m_\chi < 5.0\GeV$ for coupling to all 3 generations
of leptons.
Given the theoretical uncertainties in the standard model prediction for
the anomalous magnetic moment, we do not believe this region of parameters
is strongly preferred, but it is intriguing that this is the region where the direct detection constraints are weak.  However, to be consistent with the ATLAS search pushes one to the strong coupling region at larger $m_L$ where $\lambda > 3$ and thus these leading order results are not expected to be reliable.


\FloatBarrier

\subsection{Real Scalar and Majorana Dark Matter}
The collider bounds for the real scalar (Majorana) dark matter are identical
to the bounds for complex scalar (Dirac) dark matter.
This is because the only production mechanism is via the coupling to a photon
or $Z$, so the bounds are independent of the coupling $\la$.  

The dominant direct detection cross section arises from 2-loop diagrams
contributing to WIMP-nucleus scattering, and these are orders of magnitude
below current and projected experimental limits.
Following the discussion in \cite{Kopp:2014tsa}, Majorana dark matter has an anapole moment which normalized to the Bohr magneton is $\frac{\mathcal{A}}{\mu_N}\sim 10^{-6}~\text{fm}$ or smaller in 
the thermal relic parameter space. 
As inferred in \cite{Kopp:2014tsa}, the LUX limits are sensitive to values $\sim 10^{-5}~\text{fm}$ 
or higher.  Even with projected XENON1T sensitivity, we find that this will not place limits on this model either. 
For real scalar dark matter, the direct detection must go through two photons as shown in Fig.~\ref{fig:TwoLoopDD}, which leads to an extremely suppressed cross section below $10^{-53} \text{ cm}^2$, which will never be reached by any future experiment.

\section{Conclusions}
We have considered minimal extensions of the standard model containing
electroweak singlet dark matter with renormalizable couplings to
leptons and lepton ``partner'' particles.
These models naturally explain the observed relic abundance of dark matter
by the ``WIMP miracle.''
The strength of the coupling is fixed by requiring the correct relic
abundance, so these models are parameterized only by the masses of the
dark matter and the lepton partner particles.
This gives a simple model where direct detection and
dark matter searches at colliders can be compared under well-defined physical
minimality assumptions.

Previous studies focused on effective WIMPs coupled to colored standard
model particles, while this paper focuses on effective WIMPs coupled
to leptons.
As in the previous case, we find that collider and direct detection experiments
are remarkably complementary.
The collider limits depend only on the spin and masses of the particles,
and not whether dark matter is its own antiparticle.
The dominant constraint comes from the dilepton plus MET channel.
We find that the mono-lepton channel is less sensitive in these models.

Direct detection experiments place interesting
bounds only in the case where the dark 
matter is not its own anti-particle, {\it i.e.}~Dirac or complex scalar dark matter.
In these cases, current direct detection limits are generally stronger than
collider limits, although there is an interesting region near $m_\chi \sim m_L/2$,
where collider limits are competitive or stronger.
In this region, it is possible to discover the dark matter in both future collider
and direct detection experiments.  In particular, the next generation of direct detection experiments (e.g. XENON1T) will 
be sensitive to the bulk of parameter space.

However, if the dark matter is its own antiparticle ({\it i.e.}~Majorana or real scalar
dark matter) the direct detection bounds are well below the reach of current and next generation
experiments, while the collider bounds are unaffected.
In this case, leptophilic effective WIMPs can be discovered only in collider
experiments.

{\it Note:
As this work was being completed \Ref{Bai:2014osa} appeared,
which analyzes the same models without the relic abundance constraint
and makes  projections for lepton partner reaches at the 14~TeV LHC.}

\section*{Acknowledgements}   
SC would like to thank Graham Kribs and Yang Bai for discussions.  SC was supported in part by the Department of Energy under grant DE-SC0009945.  RE, JH, and ML were supported in part by the Department of Energy under grant DE-FG02-91ER40674.

\appendix{Appendix A: Model Details\label{sec:Appendix}}

\subsection{Dirac Dark Matter\label{diracdd}}

\paragraph{Relic Density:}
The relic density is determined by the velocity-averaged annihilation cross section $ \left\langle \si v \right\rangle$ which is commonly parametrized by the coefficients 
\[  \left\langle \si v \right\rangle \simeq a+b v^2 \]
In this model, for the annihilation cross section $\chi \chi^\dag \rightarrow \ell \ell^\dag $, $a$ and $b$ are found to be 
\beq
  a = \frac{ m_\chi^2 \sqrt{1-r} \la^4}{32 \pi  \left(m_L^2-m_\chi^2 (r-1)\right)^2}
\eeq
\beq
  \begin{split}
b =& \frac{ \la^4 m_\chi^2}{768 \pi  \left(m_L^2-m_\chi^2 (r-1)\right)^4 \sqrt{1-r}}
\big[m_L^4 \left(8-7 r+2 r^2\right)\\&+m_\chi^4 (r-1)^2 \left(-8+9 r+2 r^2\right)+2 m_L^2 m_\chi^2 \left(-12+13 r+r^2-2 r^3\right) \big]
  \end{split}
\eeq
To lowest order in $r = m_l^2 / m_{\chi}^2$, 
\beq
a \overset{r \rightarrow 0}{\simeq} \frac{ \la^4 m_\chi^2}{32\pi \left(m_L^2+m_\chi^2\right)^2  }
\eeq
\beq
b \overset{r \rightarrow 0}{\simeq} -\frac{\la^4 m_\chi^2 \left(-m_L^4+3 m_L^2 m_\chi^2+m_\chi^4\right)}{96\pi \left(m_L^2+m_\chi^2\right)^4  }
\eeq
so the cross section is not $s$-wave suppressed.

\paragraph{Direct Detection:}
One loop diagrams of the form, Fig.~\ref{fig:OneLoopDD} give rise to a charge radius, $b_\chi$,  and dipole moment, $\mu_\chi$.
\bea
\nonumber
\eql{diracbx}
b_\chi=\frac{e \lambda^2}{32\pi^2} \int_0^1 dw & \bigg[& \frac{((3w^3-3w^2+3w-2) \overline{\Delta}_1-(w-1)^3(w^2 m_\chi^2-m_l^2))}{6 \overline{\Delta}_1^2} \\
&&-m_\chi^2 \frac{w^2(2w^2-3w+1)}{3 \overline{\Delta}_1^2} -\frac{w(w-1)}{2\overline{\Delta}_1}\bigg]
\eea
and
\bea
\mu_\chi = \frac{e \lambda^2 m_\chi}{32\pi^2} \int_0^1 dw \left[-\frac{w(w-1)}{\overline{\Delta}_1}\right].
\eea
where 
\bea
  \overline{\Delta}_1 \equiv w(w-1) m_\chi^2+ (1 - w) m_l^2+ w\, m_{\tilde{l}}^2.
\eea
For the case of the electron partner, we replace $m_e$ with $40$ MeV since the minimum momentum transfer to the Xenon target at LUX is of this size, which cuts off the logarithmic divergence in the charge radius.   

These lead to a scattering cross section from the charge radius interaction,
\begin{equation}
\frac{d\sigma_{b_\chi}}{dE_R} = \frac{m_N}{2\pi v^2}Z^2 e^2 b_\chi^2 F^2[E_R]
\end{equation}
and the magnetic dipole moment interaction,
\beq\eql{diracdipole}
\frac{d\sigma_\text{DZ}}{dE_R} 
= \frac{Z^2 e^2}{4\pi E_R} \mu_\chi^2 [ 1 - \frac{E_R}{v^2}(\frac{1}{2m_N}+\frac{1}{m_\chi})]F^2[E_R]
\eeq

\subsection{Complex Scalar Dark Matter\label{complexdd}}
\paragraph{Relic Density:}
\beq
a = \frac{ \la^4 m_\chi^2 (1-r)^{3/2} r}{16 \pi  \left(m_L^2-m_\chi^2 (r-1)\right)^2}
\eeq
\beq
  \begin{split}
b =& \frac{\la^4 m_\chi^2 \sqrt{1-r}}{384 \pi  \left(m_L^2-m_\chi^2 (r-1)\right)^4}
\Big[ m_\chi^4 (r-1)^2(9 r^2-18r+8) \\ &-2m_L^2 m_\chi^2 (9r^3-31r^2+30 r-8) +m_L^4 (9r^2-2r+8) \Big]
  \end{split}
\eeq
To lowest order in $r$ we get,
\beq
a \overset{r \rightarrow 0}{\simeq} \frac{ \la^4 m_\chi^2 r}{16\pi (m_L^2+m_\chi^2)^2} 
\eeq
\beq
b \overset{r \rightarrow 0}{\simeq} \frac{\la^4 m_\chi^2}{48\pi \left(m_L^2+m_\chi^2\right)^2} 
\eeq
exhibiting the chiral suppression of $a$.  

\paragraph{Direct Detection:}
One loop diagrams of the form, Fig.~\ref{fig:OneLoopDD} give rise to a charge radius, $b_\chi$, 
\bea
b_\chi = \frac{e \lambda^2}{32\pi^2}  \int_0^1 dw\bigg[ \frac{w^3(\overline{\Delta}_1(6-4w)+w(m_\chi^2(w-1)^2+m_L^2))}{12\overline{\Delta}_1^2} - (m_L^2 \leftrightarrow m_l^2)\bigg]. \label{complexscalarbx}
\eea
where 
\bea
  \overline{\Delta}_1 \equiv w(w-1) m_\chi^2+ (1 - w) m_l^2+ w\, m_{\tilde{l}}^2
\eea
Again, for the case of the electron partner, we replace $m_e$ with $40$ MeV since the log divergence is cut off by  the minimum momentum transfer to the Xenon target.
The charge radius scattering cross section is,
\begin{equation}
\frac{d\sigma_{b_\chi}}{dE_R} = \frac{m_N}{2\pi v^2}Z^2 e^2 b_\chi^2 F^2[E_R]
\end{equation}

\subsection{Real Scalar Dark Matter\label{realscalardd}}
\paragraph{Relic Density:}
\beq
 a =\frac{ m_\chi^2 (1-r)^{3/2} r \la^4}{4 \pi  \left(m_L^2-m_\chi^2 (r-1)\right)^2} 
\eeq
\beq
b = \frac{m_\chi^2 \sqrt{1-r} r \left(9 m_L^4 r+m_\chi^4 (r-1)^2 (-16+9 r)-2 m_L^2 m_\chi^2 \left(16-25 r+9 r^2\right)\right) \la^4}{96 \pi  \left(m_L^2-m_\chi^2 (r-1)\right)^4}
\eeq
Both $a$ and $b$ vanish as $r\rightarrow 0$.  
To lowest order,
\beq
 a \overset{r \rightarrow 0}{\simeq} r\frac{ m_\chi^2 \la^4}{4\pi \left(m_L^2+m_\chi^2\right)^2 } 
 \eeq
\beq
 b\overset{r \rightarrow 0}{\simeq} -r\frac{m_\chi^4 \left(2 m_L^2+m_\chi^2\right) \la^4}{6\pi \left(m_L^2+m_\chi^2\right)^4  }
 \eeq

\paragraph{Direct Detection:}
The 2-loop contribution to the scattering cross section can be estimated using the effective operators from the relic density calculations.  After integrating out L, 
\begin{equation}
\mathcal{L}_{eff} \sim \sum\limits_l  f_l m_l  \chi^2 \bar{l}l
\end{equation}
where,
\beq
f_l =\frac{ \lambda^2}{2(m_L^2 -m_\chi^2)}.
\eeq
Ref.~\cite{Kopp:2009et} estimated the 2-loop contribution in the context of effective operators, from which we obtain
\beq
\frac{d\sigma}{dE_R} =\left (\frac{\alpha_{EM} Z}{2}\right)^4  \frac{E_R}{ v^2}\left(\frac{ 2 m_l^2\lambda^4}{3\pi(m_L^2-m_\chi^2)^2}\right)\tilde{F}(q)
\eeq
where $\tilde{F}(q)$ is the form factor for two photon scattering.

In comparison, $\frac{d\sigma}{dE_R}$  is $\mathcal{O} (10^8)$ times smaller for Real Scalar dark matter than for Dirac dark matter across the parameter space of interest, far below current experimental limits.

\subsection{Majorana Dark Matter\label{Majoranadd}}
\paragraph{Relic Density:}
\beq
a = \frac{ m_\chi^2 \sqrt{1-r} r \la^4}{32 \pi  \left(m_L^2-m_\chi^2 (r-1)\right)^2}, 
\eeq
\beq
  \begin{split}
b =& \frac{ \la^4 m_\chi^2}{768 \pi  \left(m_L^2-m_\chi^2 (r-1)\right)^4 \sqrt{1-r}}
\big[-2 m_L^2 m_\chi^2 r \left(22-35 r+13 r^2\right)\\
&+m_L^4  \left(16-26 r+13 r^2\right)+m_\chi^4 (r-1)^2 \left(16-10 r+13 r^2\right) \big]
  \end{split}
\eeq
Thus, in the massless quark limit, the $s$-wave vanishes and we have the leading order results
\beq
a \overset{r \rightarrow 0}{\simeq} \frac{ m_\chi^2 r \la^4}{32\pi(m_L^2+m_\chi^2)^2}
\eeq
\beq
    b  \overset{r \rightarrow 0}{\simeq} \la^4\frac{m_\chi^2 \left(m_L^4+m_\chi^4\right)}{48\pi \left(m_L^2+m_\chi^2\right)^4  }
\eeq

\paragraph{Direct Detection:}
Following \cite{Kopp:2014tsa}, the anapole moment for low 4-momentum transfer is given by
\beq
\mathcal{A} = -\frac{e \lambda^2}{96 \pi^2 m_\chi^2} \bigg[
          \frac{3}{2} \log \frac{\mu}{\epsilon}
            - \frac{1 + 3\mu - 3\epsilon}{\sqrt{(\mu-1-\epsilon)^2 - 4\epsilon}}
              \text{arctanh}\bigg( \frac{\sqrt{(\mu-1-\epsilon)^2 - 4\epsilon}}{\mu - 1 + \epsilon} \bigg)
        \bigg]
\eeq
where $\mu = \frac{m_L^2}{m_\chi^2}$ and $\epsilon=\frac{m_\ell^2}{m_\chi^2}$. When the 4-momentum transfer $|q^2|$ is comparable to the lepton mass as in the case of electron, the anapole moment is given by 
\beq
\mathcal{A} = -\frac{e \lambda^2}{32 \pi^2 m_\chi^2} \bigg[
                  \frac{-10 + 12 \log\frac{\sqrt{|q^2|}}{m_\chi} - (3 + 9 \mu) \log(\mu -1) -
                        (3 - 9 \mu) \log\mu}{9 (\mu - 1)}
                 \bigg]
\eeq
Scattering due to anapole scattering is much larger than the 2-loop contribution.  We compared to the LUX limits on the anapole moment derived in \cite{Kopp:2014tsa} and found that this was not a strong constraint on our parameter space even with the improvement expected for XENON1T.  

\bibliography{darkm}
\bibliographystyle{utphys}

\end{document}